\documentclass[12pt,preprint]{aastex}
\usepackage{natbib}

\begin{document}

\title{On the Dispersal of Young Stellar Hierarchies}

\author{Bruce G. Elmegreen}
\affil{IBM T. J. Watson Research Center, 1101 Kitchawan Road, Yorktown Heights, New
York 10598 USA} \email{bge@us.ibm.com}

\begin{abstract}
Hierarchical structure in young star fields has been demonstrated in a variety of ways,
including two point correlation functions (TPCFs) that are power laws for spatial
scales up to at least several hundred parsecs. As the stars age, this power law
decreases in slope until it becomes nearly flat at $\sim100$ Myr, at which point the
hierarchical structure has disappeared. The fact that the TPCF remains nearly a power
law during this time implies that the dispersal mechanism is somewhat independent of
scale. This rules out dispersal by random stellar motions at either the local gas
turbulent speed or a constant speed, because in both cases the hierarchy would
disappear at small scales first, causing the TPCF to bend over. Destruction by shear
has the right property as the shear rate in a galaxy is independent of scale for
kpc-size regions, but shear converts the hierarchy into an azimuthal stream which still
has a power-law TPCF. What does explain the observation is the overlapping of several
independent hierarchies from successive generations of star formation in the same
region. If stellar age is determined from magnitude intervals on the main sequence of a
color-magnitude diagram, or if clusters ages are grouped together logarithmically into
bins, then multiple generations will overlap more and more as the grouped populations
age, and this overlap will lower the spatial correlations between group members. Models
of these processes illustrate their relative roles in removing the appearance of young
stellar hierarchies.
\end{abstract}

\section{Introduction}

Young stars tend to be grouped together into a hierarchy of scales where smaller and
denser sub-regions are inside larger and looser regions, spanning a wide range from
star clusters and their initial subclumps at $\sim1$ parsec scales and smaller, to OB
associations at $\sim10$ parsecs, to star complexes at $\sim100$ parsecs, to flocculent
spiral arms on the largest scale \citep[see review in][]{elmegreen10}.

Hierarchical structure on large scales was observed for whole OB associations in the
LMC \citep{feitzinger84}, HII regions in 19 galaxies \citep{feitzinger87} and 93
galaxies \citep{sanchez08}, stellar groupings determined from near-neighbor path
linkage in M31 \citep{battinelli96}, NGC 300 \citep{pietrzynski01}, and in seven other
galaxies \citep{bresolin98}, and star complexes using flux contours in M51
\citep{bastian05} and in nine other galaxies \citep{gusev02}. It was also observed
using stellar density contours in the LMC \citep{maragoudaki98} and M33
\citep{ivanov05}, and box counting techniques in 10 galaxies \citep{elmegreen01a}, 14
galaxies \citep{elmegreen14} and at high resolution in NGC 628 \citep{elmegreen06}. It
was observed with Fourier transform power spectra of optical light in six galaxies
\citep{elmegreen03a}, optical and H$\alpha$ light in nine dwarf irregulars
\citep{willett05}, and optical light at high resolution in M33 \citep{elmegreen03b}.
These studies looked primarily at the positions of star-forming regions and did not
consider the time evolution of this structure.

Hierarchical structure among individual stars gives some information about evolution
because a star's position on the main sequence of a color-magnitude diagram gives an
upper limit to its age. Stellar hierarchies have been observed in M33 and NGC 6822
using minimum spanning trees \citep{bastian07,gouliermis10}, in the LMC
\citep{harris99,bastian09} and SMC \citep{gieles08} using several methods including the
$Q$ parameter from \cite{cartwright04} and the two point correlation function (TPCF),
in NGC 6503 and NGC 1566 with the TPCF \citep{gouliermis15,gouliermis17}, in four
\citep{odekon06} and six \citep{bastian11} dwarf irregular galaxies using the TPCF or
$Q$ parameter, and in the LMC, SMC, M31 and M33 over at least two orders of magnitude
in scale with the TPCF \citep{odekon08}.

Individual star forming regions in the SMC \citep{gouliermis12} and Milky Way
\citep{gomez93,larson95,simon97,gomez98,bate98,johnstone00,testi00,smith05} were also
shown to be hierarchical on smaller scales using the TPCF. Sub-structure has been noted
in resolved OB associations and clusters too
\citep[e.g.,][]{heydari01,kumar04,dahm05,gutermuth08}. Embedded stars tend to follow
the fractal structure of the clouds in which they form
\citep{sanchez07,fernandes12,gregorio15}. Stellar multiplicity
\citep[e.g.,][]{brandeker03} could be the small scale limit of hierarchical structure.

Generally, these studies find that hierarchical structure in stars disappears gradually
over time, taking around 100 Myr for scales between $\sim1$ pc and several hundred
parsecs. Substructure in open clusters disappears over time too
\citep{sanchez09,sanchez10}. Observations of clustering in red supergiants and Cepheid
variables \citep{payne74,efremov78,elmegreen96}, for which individual ages may be
derived, were reviewed by \cite{efremov95}; usually there are not enough of these stars
in any one region to see levels of substructure.

Position and time correlations among clusters give the most detailed picture of
evolution because each cluster has an age. \cite{zhang01} used the TPCF to show
hierarchical structure up to $\sim0.7$ kpc and 160 Myr for clusters in the Antennae
galaxy; there was no clear variation with age in that range although massive young
stars less than 10 Myr old had less correlation than the clusters. \cite{scheepmaker09}
measured the TPCF in M51 for clusters in three logarithmic age bins up to 400 Myr and
found that older clusters were slightly less correlated than younger clusters on scales
larger than 160 pc.  \cite{grasha15} showed a strong two-point correlation from the
limit of resolution at 0.2 arcsec (9.6 pc) up to 100 arcsec (4800 pc) for Class 3
clusters (which are young and possibly unbound) in NGC 628.  \cite{grasha17a} found
similar correlations for all resolved separations below $\sim1$ kpc for young stellar
groupings in 5 other galaxies, and showed that this upper limit increases with galaxy
size, ranging from several hundred parsecs in NGC 7793 and NGC 3738 to several thousand
parsecs in NGC 1566 and NGC 628.

The decrease in spatial correlation between clusters with increasing age suggests not
only that older clusters are more randomly positioned than younger clusters, but also
that clusters close to each other in space tend to be younger, which means that they
have more similar ages than clusters that are far from each other. This correlation is
such that the age difference, $\Delta T$, between sub-regions in a hierarchy of star
formation increases approximately as their spatial separation, $\Delta S$, to some
power. In the LMC, $\Delta T= 2.9\Delta S ^{0.33}$ between $\Delta T\sim8$ Myr at
$\Delta S\sim20$ pc and $\Delta T\sim30$ Myr at $\Delta S\sim1$ kpc \citep{efremov98}.
In the local Milky Way, $\Delta T = 10.6\Delta S^{0.16}$ from $\sim20$ Myr at $\sim50$
pc to $\sim30$ Myr at $\sim600$ pc \citep{marcos09}. In 8 other galaxies, $\Delta T
\sim 5.0\Delta S^{0.38}$ from an average age difference of $\sim8.6$ Myr at $\sim4.2$
pc to $\sim55$ Myr at $\sim550$ pc \citep{grasha17b}. These power laws are reminiscent
of the correlation between the crossing time inside a molecular cloud, $T_{\rm cross}$,
and the cloud radius, $R$. The crossing time is the ratio of twice the cloud radius,
$2R$, to the internal velocity dispersion, $\sigma$. For the Milky Way,
$\sigma=0.48(R/{\rm pc})^{0.63\pm0.30}$ km s$^{-1}$ for $R=1$ to 200 pc
\citep{miville17}, so $T_{\rm cross}=4.1R^{0.37}$ Myr for molecular clouds. This
implies $\Delta T\sim T_{\rm cross}$ for young regions \citep[see also][]{elmegreen00}.

All of these correlations for young stars and clusters disappear as the stars and
clusters age. This seems at first a reasonable expectation because of random motions.
However, random motions produce an expectation that is not observed, namely that
smaller scales should lose their correlated structures before larger scales. This is
because in a given scale, $\lambda$, random mixing removes substructures smaller than
the mixing time, $\tau=\lambda/\sigma$, for velocity dispersion $\sigma$, and it does
not significantly remove structures larger than this. Thus the power law TPCF should
have a break and turn over on scales smaller than $\tau\sigma$, and this turn-over
distance should increase linearly with time if $\sigma$ is independent of scale, and as
the second power of time if $\sigma\propto \lambda^{0.5}$.  Observations of the time
dependence of stellar correlations \citep{gieles08,gouliermis15,gouliermis17} show that
the TPCF remains a power law, however. The TPCF amplitude decreases uniformly over time
while preserving the power law \citep[see also][]{bastian09,bastian11}, which implies
that every scale gets washed out simultaneously.

Another process that can remove correlated structure is shear. Indeed, \cite{grasha17b}
found that the ratio of the maximum spatial separation in the time-space correlation
for young clusters to the cluster age difference is always about equal to the relative
velocity from shear on this maximum scale. That is, the maximum age difference for the
observed correlation is always about the inverse of the Oort $A$ shear parameter for
the 8 galaxies they studied.  Shear has the attractive property that the shear rate is
independent of scale, since $A$ does not vary much on scales below a kiloparsec or so.
Thus shear could decrease the amplitude of the TPCF uniformly with time, preserving the
power law. However, shear just stretches the initial structure in the azimuthal
direction, removing the radial component of the correlation and not the azimuthal
component, so the decrease in TPCF cannot go to zero amplitude, only to the correlation
amplitude expected for a 1-dimensional fractal. Shear plus random motion also has the
wrong property because the time-dependent turn-over at small scales would still be
present from the random motions.

A third possibility is that the correlated structure in aging star fields disappears
because new regions of star formation with new correlations occur on top of the old
regions. This {\it superposition} removes the correlation of the summed distribution in
a scale-independent way over time, decreasing the amplitude of the total TPCF to zero
if the overlap is large enough.  Selecting stellar age groups from the color-magnitude
diagram, which has been the usual procedure for some of the above papers, can have this
superposition effect because low mass stars can be both old according to their
main-sequence turn off positions, and young as members of more recently formed groups
\citep[e.g.,][]{bastian11}. Selecting stellar ages in logarithmic intervals also has
this effect because the time interval for the selection increases with age, and so the
number of separate star-forming episodes in each region increases with age too. TPCFs
for clusters can in principle get around this blending effect because the ages are
known and different age groups can be examined separately. However, the number of
clusters at small separations is usually low at the older ages, so any predicted
small-scale turnover in the TPCF, e.g., from random motions, might be difficult to see.

In what follows, Section \ref{Setup} discusses the model, Sections \ref{diff} and
\ref{shear} show the effects of random stellar motions and shear, and Section
\ref{super} models the superposition of stellar populations, without and with the
inclusion of random motions and shear. The conclusions are in Section \ref{conc}.

\section{A Model for Hierarchical Distributions}
\subsection{Setup}
\label{Setup}

The effects of random motion, shear, and superposition on the TPCF are studied here in
the shearing sheet approximation for a piece of a galaxy.  The initial distribution of
young stars is taken to be a random fractal on a plane made by selecting, in a
hierarchy of steps, $N_{\rm f}$ subcells inside each cell with a decrease in cell size
equal to a factor of $f=1/2$ at each step. The fractal dimension in this case is $-\log
N_{\rm f}/\log f$.

In a typical model, the sheet consists of 512 by 512 lowest-level cells with 8
($1+\log_2 512/4$) sub-levels of hierarchy.  The highest level has 16 cells, i.e.,
$4\times4$ with side lengths of 128 cell-lengths each, and these 16 cells are
independent. Lower levels have $2^i\times2^i$ cells of $512/2^i$ cell-lengths each, for
$i=3$ to 9. Starting with the $4\times4$ cells, a number of them is chosen to contain
young stars in their sublevels. This number has an average value of $p\times4\times4$
where $p=2^{D_{\rm f}-2}$ is the probability of choice determined by the fractal
dimension $D_{\rm f}$. For example, if the fractal dimension is $D_{\rm f}=2$, then
$p=1$ and all cells are chosen, filling in the grid completely. We pick $D_{\rm f}=1.3$
here in analogy with interstellar clouds \citep{elmegreen96b}, and then $p=0.62$.  The
cells are chosen by cycling through each one and picking a random number uniformly
distributed between 0 and 1, and then determining if that random number is larger or
smaller than $p$. If it is smaller, than we use that cell for subsequent division into
subcells, and if it is larger, we ignore that cell during further steps. We next cycle
through each of the chosen cells from this $4\times4$ level to pick subcells in the
next lower level. Again we pick a random number for each possible subcell and compare
it to $p$. After we have this next list of chosen cells, we cycle through each of
those, picking sub-subcells with the same probability $p$. Eventually we get to the
lowest cell in the grid, which cannot be subdivided. We consider each of the chosen
lowest-level cells to contain one object whose TPCF is to be measured, i.e., a star or
a cluster. For convenience, the size of the overall grid is normalized to 1, which
means that each cell has a size of $1/2^i$ for $i=2$ on the largest scale to $i=9$ on
the smallest scale.

Once the cells with stars are chosen, the TPCF is determined by cycling through each
star (called the ``first star'' here) and counting the number of other stars (the
``second star'') within certain intervals of distance. These intervals are equally
spaced in units of the log of the distance. To avoid boundary effects, we pick only
first stars between the limits of 0.25 and 0.75 times the grid coordinates, and
consider distances between these first stars and the second stars only up to 0.25 of
the grid size. Thus the range of possible distances between stars is 0.25 of the grid
size, or 128 cell units. This is the maximum spatial range of the power law in the
resulting TPCF. For a uniform distribution of stars in the plane, the number of pairs
increases with distance to the power 2 for logarithmic distance intervals. To find the
excess correlation above this uniform value, we make a histogram of log-distances by
summing $(0.25/D)^2$ for each pair with distance $D$ (recall that the longest distance
is 0.25). We confirmed that for a uniform distribution of stars, the histogram is flat.
This procedure gives the initial TPCF, i.e., at the beginning of the time evolution
before random motions, shear, or superposition are included.

\subsection{Random Motions}
\label{diff}

To simulate the effects of stellar motions with a random turbulent velocity, the stars
in each hierarchical level $i$ from 2 to 9 were moved for a distance given by a
simulated velocity dispersion, $\sigma_{\rm i}$, that depends on the level. Recall that
the cells have sizes $1/2^i$, so we choose $\sigma_{\rm i}$ to scale with the square
root of the cell size. Thus all of the stars in each cell of level $i$ move because of
initial cloud turbulence at that level, and they move for a distance in the $x$
direction equal to $\sigma_0/2^{i/2}$ times a random variable between $-0.5$ and $0.5$.
Similarly, they move for a distance in the $y$ direction equal to this same fiducial
value times another random variable between $-0.5$ and $0.5$. This motion occurs
hierarchically, which means that all of the stars in each large cell, with $i=2$, move
together, although each large cell moves differently. At the same time, all of these
stars move again for a shorter distance in batches according to their groupings in the
next-smaller cells, which have $i=3$. The same stars move again with smaller distances
grouped into the $i=4$ cell, and so on until the $i=9$ cell, which contains only one
star. These single stars finally move once again over the smallest relative distance.
Thus each star has 8 contributions to its total motion, each in a random direction, and
one for each of the 8 cells it is in with one cell per hierarchical level from $i=2$ to
9. These 8 contributions reflect the concept that turbulence consists of smaller scale
motions inside larger scale motions, and that stars are born with a motion equal to
that of the cloud in which they form.

\begin{figure}
\epsscale{.7}
\plotone{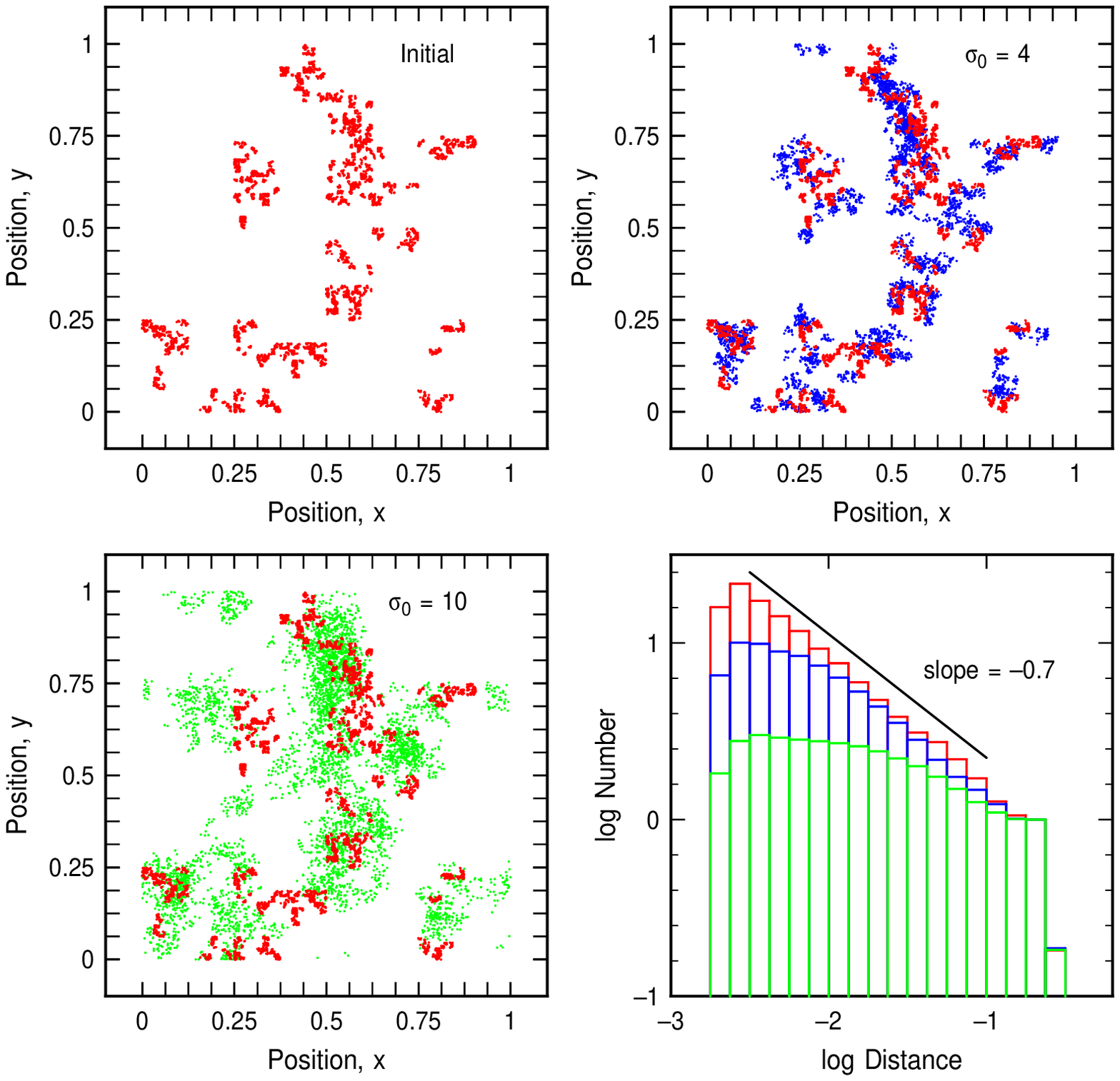}
\caption{The positions of stars in a fractal model are shown. In the upper left they have their initial state,
in the upper right they have been given random displacements
(blue points) with amplitude $\sigma_0=4$ (see text), and in the
lower left they have been given larger random displacements (green points) with $\sigma_0=10$. The initial state is repeated as red
points for comparison. The histograms in the lower right are the TPCFs with colors corresponding
to the points; each TPCF is an average of 10 random trials with the same
parameters but different random initial conditions.
The initial TPCF (red) is a power law with a slope of $-0.7$ as expected for the
fractal dimension on a plane, $D=1.3$. As the stars move randomly, increasing their
excursions over time (from red to blue to green point distributions),
the power law bends over first at small scales and then continues to bend over at larger scales.
}\label{shearsheet3}
\end{figure}

Figure \ref{shearsheet3} shows the TPCF and its evolution with random motions. The
initial distribution for one of ten trial fractals is shown by the red points in each
panel. The initial TPCF is shown by a red histogram in the lower right; this is the
average of 10 histograms made from different random initial conditions. The TPCF is a
power law with a slope $\gamma=-0.7$, as expected for the initial fractal distribution,
which has a fractal dimension of $D_{\rm f}=1.3$ ($\gamma=D_{\rm f}-2$ in two
dimensions). The distribution after random motions with amplitude $\sigma_0=4$ is shown
by the blue points in the upper right and the blue histogram (also an average of 10
histograms). The distribution with $\sigma_0=10$ is shown by the green points in the
lower left and the green average histogram.  The main effect of random motions on the
TPCF is a turnover at small scale, with increasing scale for the turnover at larger
random motions. Larger random excursions in our model correspond to older stars, as
they have moved further from their origins (ignoring epicycles).

\subsection{Shear}
\label{shear}

To simulate the effects of shear, we move each star for a distance in the $x$ direction
(horizontal in the figures, and azimuthal in the galaxy) given by $\Delta x = A(y-0.5)$
(where $y$ is vertical in the figures and radial in the galaxy). This motion is
independent of the level of the hierarchy and it assumes that each star forms at the
same time.

\begin{figure}
\epsscale{.9}
\plotone{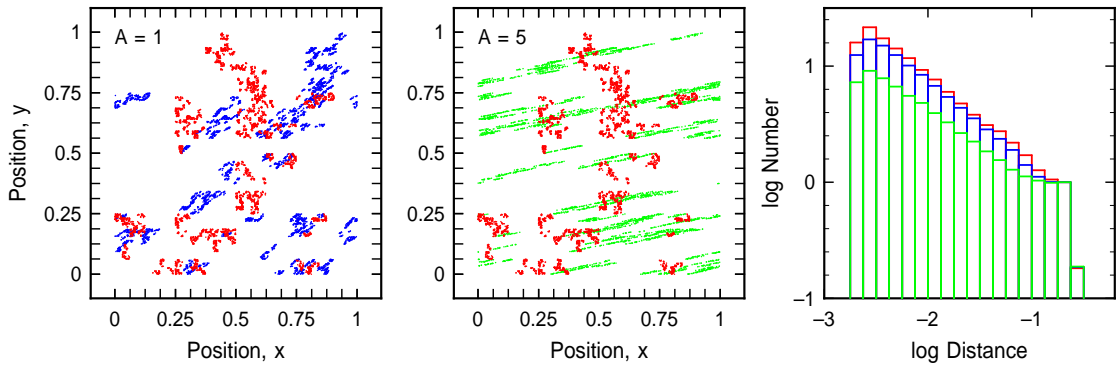}
\caption{The positions of stars in the fractal model with shear increasing from red points
(initial state) to blue to green, as measured by the shear variable $A=1$ and $A=5$ (see text).
The TPCFs are shown with the same colors as the points (each is an average of 10 trials).
The TPCF flattens first at
large scales because shear moves stars apart most on large scales, but the overall flattening
is limited by the remaining correlation in the azimuthal direction (horizontal in the figures).
}\label{shearsheet4b}
\end{figure}

Figure \ref{shearsheet4b} shows the result. The red points in the left and middle
panels and the red histogram are the same initial positions and TPCF as in Figure
\ref{shearsheet3}. The blue points on the left correspond to $A=1$ and are compared to
the red points in the same panel. The green points in the middle panel are for $A=5$.
Blue and green histograms show the corresponding TPCFs. The TPCFs are averages of 10
trials with random initial conditions. Note that $A=1$ corresponds to a horizontal
motion of 0.5 times the total grid size in the rightward direction at the top of the
grid and the same motion to the left at the bottom of the grid. The transformation of
red points to blue points on the left shows the patches moving toward a 45 degree
angle. Points that move off the grid to the left and right are wrapped around to the
other side. For the larger shear value, $A=5$ in the middle panel, points near the top
and bottom of the grid have been wrapped around many times.

The histograms decrease in amplitude with increasing shear, but they do not change
significantly in slope and the changes are not very fast. This is because local regions
do not shear much within themselves but tend to lose correlation first with the more
distant regions. Thus the power law flattens most on large scales and it does not
change much on small scales. Also, stars that are initially correlated in the azimuthal
direction do not lose their correlation because their relative positions do not change
with shear.

\subsection{Superposition}
\label{super}

After stars form in a region, the gas that formed them moves around and later forms new
stars. Presumably this is a continuous process for a steady star formation rate, and
presumably each new generation is somewhat independent of the old one. Here we
determine the TPCF for stars made in several generations. We first show the result
without random motions or shear, and then we add these additional effects.

\begin{figure}
\epsscale{.7}
\plotone{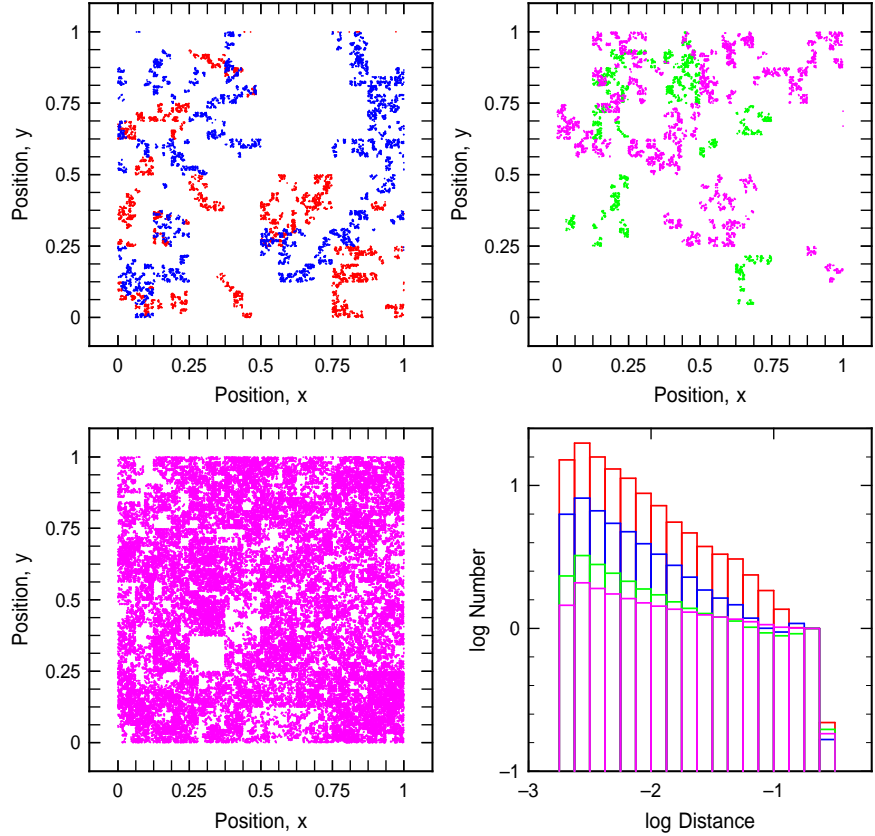}
\caption{The positions of stars in a superposition of fractal models, with no random motions
or shear. The top left panel shows the most recent population of stars (the 20th generation)
as red points and the next most recent as
blue points, both in their original positions in the absence of random motions and shear. The
red histogram is the TPCF for the one case with red points, and the blue histogram is the TPCF for the
sum of the red and blue points, taken as a composite spatial distribution. This blue TPCF shows
a flattening at large scales because the two populations overlap most on these scales, but the
power law is about the same as for a single population on small scales because each
population is still directly visible on small scales. In the upper right, the magenta points are
the first population out of 20 to form, and the green points are the 11th generation.
These points also have their initial positions because in this figure there are no
random motions nor shear. The lower left panel shows all of the 20 generations superposed.
The green histogram is from the superposition of the first 11 generations to form and the
magenta histogram is from the superposition of all of them.
}\label{shearsheet7}
\end{figure}

Figure \ref{shearsheet7} shows the result of superposition alone for 20 generations.
The top left shows the positions of stars in the 20th (last to form) and 19th (second
last to form) generations as red and blue points. Without random motion or shear, they
have not moved. The red histogram in the lower right shows the TPCF for this 20th
generation; it is a power law because the stars have their original fractal
distribution. The blue histogram in the lower right is the TPCF for the sum of the two
generations, i.e., the red and blue points together. The TPCF is still the same power
law on small and intermediate scales because each generation still has structure on
these scales and there is little overlap between the two. The TPCF for the
superposition decreases on large scales, however, because there is no correlation
between stars in the 20th generation and stars in the 19th generation, and these
intergenerational spacings tend to have large scales when only two generations are
involved.

The top right of Figure \ref{shearsheet7} shows star positions for the 1st (first to
form) and 11th generations in magenta and green, respectively. These distribution also
appear with their original fractal forms because there is no random motion and shear.
The green and magenta histograms in the lower right are from a superposition of the
first 11 generations and all 20 generations, respectively. With more generations added,
the large-scale flattening of the TPCF spreads to smaller scales as the uncorrelated
overlapping regions become more dense. The lower left panel shows all 20 generations
superposed; this is what made the magenta histogram.

The effects of superposition with random motions are shown in Figure \ref{shearsheet5}.
The random distance increases with time, so in a stack of 20 generations the oldest
stars ($j=1$ generation) have moved the most and the youngest ($j=20$ generation) not
at all. A motion rate corresponding to $\sigma_0=10$ after 20 generations is used
(i.e., $\sigma_0=(N-j)/2$ for the {\it j}-th generation out of $N=20$). The colors in
Figure \ref{shearsheet5} represent the same generations as in Figure \ref{shearsheet7}.
In the top left, the most recent generations to form (20th and 19th) have little or no
random motions, and their histograms (red and blue) are nearly power laws. In the top
right panel, the 1st generation (magenta) is very dispersed and the 11th generation
(green) less so. Their TPCFs are in the lower right with the same colors.

Figure \ref{shearsheet6} shows superposition and random motions as in Figure
\ref{shearsheet5}, but now also with shear given by $A=(N-j)/20$, which is about the
same amount of shear for the oldest stars ($j=1$) as in the left panel of Figure
\ref{shearsheet4b}, where $A=1$. The effects of shear can be seen in the stellar
distributions, especially for the magenta points ($j=1$) in the upper right panel, but
it does not change the TPCF much compared to the case of superposition plus random
motion without shear.

\section{Conclusions}
\label{conc}

Several processes in galaxy disks cause young stars and clusters to decrease their two
point correlation over time. Random stellar motions acquired at birth can do this,
producing a bend in the TPCF at small scales where the random motions are relatively
fast for their length scale. Because observations by \cite{bastian11},
\cite{gouliermis15}, \cite{gouliermis17} and others do not show such a bend, random
motion alone is probably not the only process at work. The TPCF for clusters in
\cite{grasha15,grasha17a} also shows no bend at small scales, although the uncertainty
is large because the number of clusters is small.

Shear would seem to remove correlations in the positions of young stars, and
\cite{grasha17b} show that the average shear velocity in a region corresponds to the
ratio of the maximum correlated length to the maximum cluster age difference at that
length. However shear in our models does not appear to affect the TPCF very much. Shear
causes the more distant stars and clusters to move for a larger relative distance than
the nearby stars and clusters, so it flattens the TPCF on large scales first. However,
it takes a lot of shear to change the TPCF noticeably, and the correlations in the
azimuthal direction are not changed at all.

\begin{figure}
\epsscale{.7}
\plotone{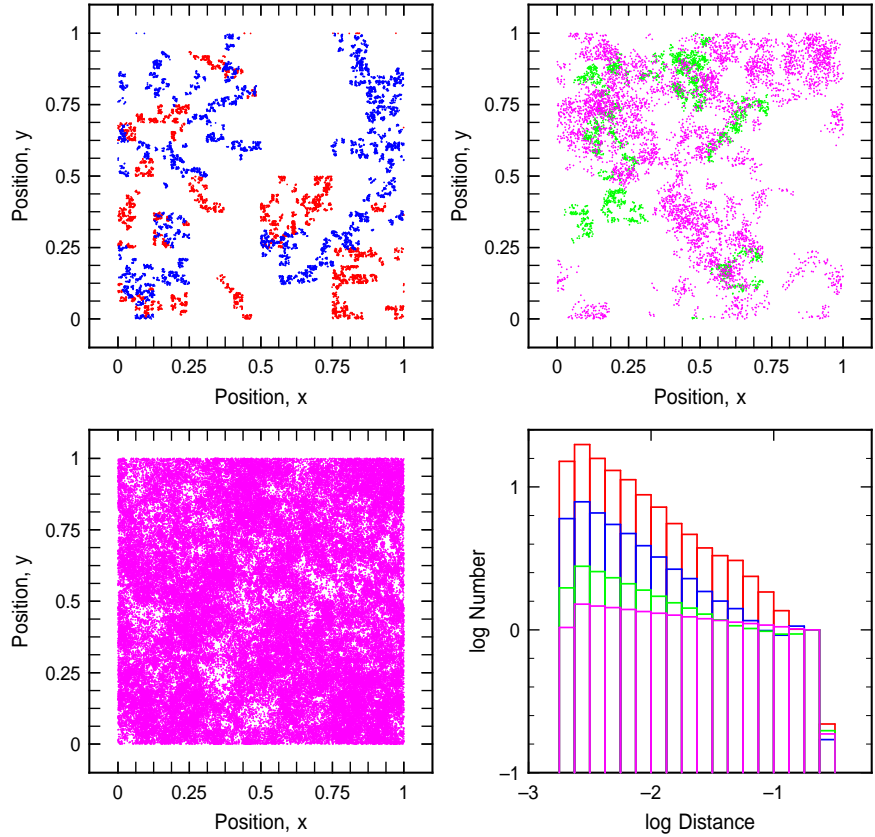}
\caption{The positions of stars in a superposition of fractal models,
this time with random motions but still
no shear. The points have the same colors as in Figure 3 but now the effect of aging and
increased dispersal are evident. The youngest population (20th generation) is red, the next
youngest is blue, the 10th youngest (the 11th to form) is green and the oldest, or 1st generation,
is magenta. The histograms flatten first at large scales, but random motions
supplement this flattening
at small scales and the result is a steady decrease in the slope of the power law.
}\label{shearsheet5}
\end{figure}

\begin{figure}
\epsscale{.7}
\plotone{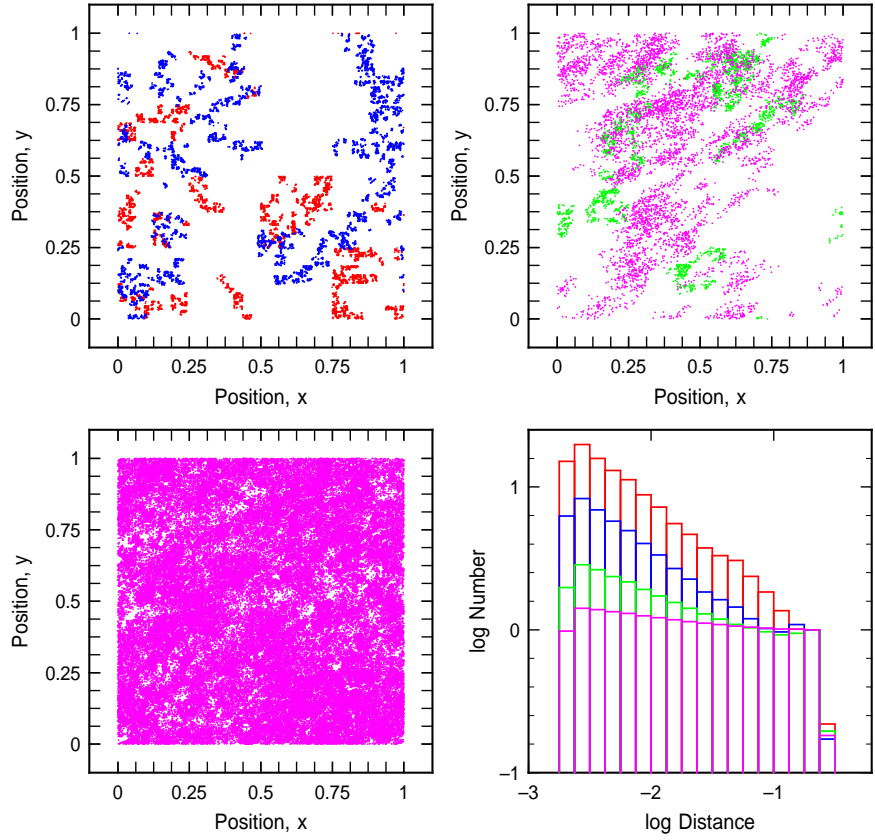}
\caption{The positions of stars in a superposition of fractal models,
this time with both random motions and shear.
The points have the same colors as in Figures 3 and 4. Shear has relatively
little additional affect on the TPCFs.
}\label{shearsheet6}
\end{figure}

Superposition of multiple generations seems to affect the TPCF most, especially when
combined with random motion (Fig. \ref{shearsheet5}). Shear can be present also, but it
does not contribute much more (Fig. \ref{shearsheet6}). This result applies when stars
with a wide range of ages are included in the TPCF, because then the younger stars are
likely to have formed in a different hierarchy of interstellar gas than the older
stars. Continuous star formation should have this property, so the appearance of a
power law TPCF with a shallow slope may indicate that the objects included in that TPCF
have been forming continuously for the duration of the sample. Star formation that
occurred in the recent past with no younger stars superposed should have a power-law
TPCF with a bend at small scales from random stellar motion. Thus, there should be a
relationship between the history of star formation and the shape of the TPCF when the
crossing time for random stellar motions increases with scale.

An exception to this result occurs if the stars are in a gravitationally bound cluster
with a radial density profile. Then the TPCF can remain a power law on small scales
even though the stars are mixed and have lost their initial hierarchy. The TPCF bends
over only when random stellar motions move stars away from each other.

In summary, the model predicts that a bend should appear on small scales in the TPCF
for young objects that are in a small age interval so that they formed nearly
simultaneously. The bend is from random stellar motions or random cluster motions in an
initially turbulent interstellar gas. The bend may not appear for objects in a large
age interval, which includes stars chosen from their appearance on the main sequence of
a color-magnitude diagram. Stellar magnitude gives only an upper limit to the stellar
age, which is the turn-off age for that star, so a range of ages is possible among main
sequence stars with the same magnitude. When there is a range of ages like this, then
the superposition of a power-law TPCF from each age interval will diminish the
correlation and flatten the function for the whole sample, starting with the largest
scales first and then moving toward smaller scales as more independent star formation
epochs are added together and the overlap becomes denser.

\clearpage

\clearpage

\clearpage

\clearpage

\clearpage

\end{document}